\documentclass[prd, 11pt, a4paper, showpacs, floatfix, groupedaddress, notitlepage, dvipsnames, reprint, twosided, onecolumn, preprintnumbers]{revtex4-1}

\usepackage{geometry}
\usepackage[utf8x]{inputenc}
\usepackage{graphicx}
\usepackage{bm, amssymb, amsmath, amsfonts}
\usepackage{multirow}

\usepackage{hyperref}
\hypersetup{
colorlinks = true, 	    % false: boxed links; true: colored links
linkcolor = magenta,	% color of internal links
citecolor = blue,		% color of links to bibliography
}
\usepackage[linesnumbered,ruled,lined,boxed]{algorithm2e}
\usepackage[noend]{algpseudocode}

\SetAlCapSty{xAlCapSty}

\newcommand{\atcp}[1]{\tcp*[r]{\parbox[t]{.575\linewidth}{#1\hfill}}}
\SetCommentSty{mycommfont}

\SetNlSty{mynlfont}{}{} 

\RestyleAlgo{algoruled}

\usepackage{caption}
\usepackage{subcaption}
\def \PyCBC	    {\texttt{PyCBC}}
\def \GstLAL	{\texttt{GstLAL}}
\def \GW        {\texttt{GW}}
\def \CNN       {\texttt{CNN}}
\def \NSBH     {\texttt{NSBH}}
\def \BBH       {\texttt{BBH}}
\def \BNS      {\texttt{BNS}}
\def \SNR     {\texttt{SNR}}
\def \CBC     {\texttt{CBC}}
\def \PSD     {\texttt{PSD}}
\def \BH        {\texttt{BH}}
\def \NS        {\texttt{NS}}
\def \LIGO      {\texttt{LIGO}}

\def \KAGRA     {\texttt{KAGRA}}
\def \LVK       {\texttt{LVK}}
\def \IMRPhenomD    {\texttt{IMRPhenomD}}
\def \ROC       {\texttt{ROC}}
\def \GPU       {\texttt{GPU}}
\def \Hz        {\texttt{Hz}}

\setcitestyle{numbers,square}

\begin{document}

\preprint{LIGO DCC number \color{blue}{\bf LIGO-P2100306}}

\title{
Employing Deep Learning for Detection of Gravitational Waves from Compact Binary Coalescences
}

\author{Chetan~Verma}	\email[]{chetanverma.phd@iar.ac.in}
\affiliation{\mbox{Institute of Advanced Research, Gandhinagar, Gujarat, India}}
\author{Amit~Reza}	    \email[]{a.reza@nikhef.nl}
\affiliation{\mbox{Nikhef, Science Park, 1098 XG Amsterdam, The Netherlands}}
\affiliation{Institute for Gravitational and Subatomic Physics (GRASP), Utrecht University, Princetonplein 1, 3584 CC Utrecht, The Netherlands}
\author{Dilip Krishnaswamy} %\email[]{dilip@ieee.org}
\affiliation{\mbox{Indian Institute of Technology Delhi, Delhi 110016, India}}
\author{Sarah Caudill}      %\email[]{sarah.caudill@nikhef.nl}
\affiliation{\mbox{Nikhef, Science Park, 1098 XG Amsterdam, The Netherlands}}
\affiliation{Institute for Gravitational and Subatomic Physics (GRASP), Utrecht University, Princetonplein 1, 3584 CC Utrecht, The Netherlands}
\author{Gurudatt~Gaur}	%\email[]{gurudatt.e11103@cumail.in}
\affiliation{\mbox{Chandigarh University NH-95 Chandigarh-Ludhiana Highway, Mohali, Punjab (INDIA)}}
%\affiliation{
%\mbox{Institute of Advanced Research, Gandhinagar, Gujarat, India}
%\mbox{Nikhef, Science Park, 1098 XG Amsterdam, The Netherlands} \\
%\mbox{Indian Institute of Technology Delhi, Delhi 110016, India} \\
%\mbox{Nikhef, Science Park, 1098 XG Amsterdam, The Netherlands} \\
%\mbox{Chandigarh University NH-95 Chandigarh-Ludhiana Highway, Mohali, Punjab (INDIA)}}

\begin{abstract}
The matched filtering paradigm is the mainstay of gravitational wave ($\GW$) searches from astrophysical coalescing compact binaries (e.g., binary black holes, binary neutron star, neutron star, and black hole pair).
The compact binary coalescence ($\CBC$) search pipelines (e.g. $\GstLAL$, $\PyCBC$) perform the matched filter between the $\GW$ detector's data and a large set of analytical waveforms. However, the computational cost of performing matched filter is very high as the required number of the analytical waveforms is also high. Recently, various deep learning-based methods have been deployed to identify a $\GW$ signal in the detector output as an alternative to computationally expensive matched filtering techniques. In past work, the researchers have considered the detection of $\GW$ signal mainly as a classification problem, in which they train the deep learning-based architecture by considering the noise and the $\GW$ signal as two different classes. However, in this work, for the first time, we have combined the Convolutional Neural Network ($\CNN$) and matched filter methods to reduce the computational cost of the search by reducing the number of matched filtering operations. We have implemented the $\CNN$ based architecture not only for classification of the signal but also to identify the location of the signal in the intrinsic parameter (e.g., mass, spin) space. Identifying the location (patch) in which the detected signal lies enables us to perform the matched filter operations between the data and the analytical waveforms generated for the smaller region of the parameter space (patch) only - thereby reducing the computational cost of the search. We demonstrate our method for two-dimensional parameter space (i.e., for non-spinning case, in component mass space $(m_{1}, m_{2})$ only) for stellar to high mass binary black hole ($\BBH$) systems. In particular, we are able to classify between pure noise and noisy $\BBH$ signals with $99\%$ accuracy. Further, the detected signals have been sub-classified into patches in mass components with an average accuracy $\geq 97\%$. \\

\vspace{12pt}
\end{abstract}

\maketitle

\section{Introduction}
\label{Sec:Intro}

The detection of gravitational waves ($\GW$) has brought us a new way of studying the properties of astrophysical objects such as black holes ($\BH$), neutron stars ($\NS$), and their formation channels via their mergers. {\color{black}{The first $\GW$ signal has been detected in $2015$ \cite{abbott2016gw150914}. The detected signal was emitted from a  
system of binary black holes ($\BBH$).} After first detection,} so far,  more than $50$ $\GW$ events \cite{abbott2019gwtc, Abbott:2020niy} have been observed by the global network of $\GW$ observatories $\LIGO$-Virgo-$\KAGRA$ ($\LVK)$ \cite{aLIGO, accadia2011status, Akutsu:2020his}. The $\GW$ catalog includes the signals from binary black holes ($\BBH$), a binary neutron star ($\BNS$), and neutron star-black hole ($\NSBH$) merger events.

The detection of $\GW$s is highly dependent on a computationally expensive technique, called \textit{matched-filtering}\cite{helstrom2013statistical, Findchirp}. In this technique, the data obtained from the global network of $\GW$ detectors ($\LVK)$ is cross-correlated against a set of analytical waveforms corresponding to a fixed set of points in the intrinsic parameter space (e.g., mass components, spin components), known as template points or template bank \cite{Satya, owen1996search, harry2009stochastic}. The triggers are recorded whenever the correlation exceeds the set threshold. Since the parameters of the detected $\GW$ signal are not known a priori,  we need a large set of waveform templates  adequately covering the parameter space. The template banks are generally constructed based on two different classes of template placement algorithms, such as Stochastic \cite{harry2009stochastic}, and Geometrical \cite{owen1996search} template placement algorithms.

Matched filtering is an optimal algorithm to search an unknown signal in the presence of Gaussian noise. Mathematically it can be expressed as follows:
%%%%---------------------------------------------------------------
\begin{equation}
\langle s(t), h(t) \rangle = 4 \mathbb{Re} \, \int_{0}^{\infty} {\frac{\tilde{s}(f) \, \tilde{h}^{*}(f)}{S_{n}(f) } \, df } \, , 
\label{Eq:SNR}
\end{equation}
where $S_{n}(f)$ is defined as the one-sided power spectral density ($\PSD$) and $\mathbb{Re}$ denotes the real part. The square root of Eq.\ref{Eq:SNR} is known as Signal to Noise Ratio ($\SNR$). The detector output $s(t)$ sometimes can contain actual $\GW$ signal or sometimes pure noise. $h(t)$ represents an analytical waveform based on a specific point in the template bank. For the computation of the $\SNR$, the norm of each template waveform needs to be calculated, using systems at a distance of $1$ Mpc, which is mathematically defined as $\sigma^{2} = \langle h(t), h(t) \rangle$. Then the matched filter output, i.e., $\SNR$ can be defined as follows: 

\begin{equation}
\rho(t) = \frac{\langle s(t), h(t) \rangle}{\sigma}
\end{equation}
%%%%--------------------------------------------------------------------
It provides a measure of the degree to which a given template waveform matches the measured $\GW$ signal present in the data. Using the definition given in the Eq.\ref{Eq:SNR}, we can define the power of each template waveform in terms of optimal-$\SNR$ ($\rho_{\text{opt}}$) as follows:
%%%%------------------------------------------------------------------
\begin{equation}
\rho_{\text{opt}} = \sqrt{4\int_{f_{min}}^{f_{high}}\frac{\lvert\tilde{h}(f)\rvert^{2}}{S_{n}(f)}df}
\end{equation}

%%%%-------------------------------------------------------------------
For an effective search, these templates are placed very densely in the parameter space. Since the total number of matched filter operations is directly proportional to the number of template waveforms, this gives rise to the latency in the $\GW$ search. For example, the computational cost of matched filtering per template is $N_{s} \log N_{s}$ floating-point operations, where $N_{s}$ is the number of samples for a template waveform, that can be computed based on chosen sampling frequency ($f_{s}$) and the duration of the template waveform ($T$) as $N_{s} = f_{s} \times T$. In general, $\LVK$ uses an aligned spin template bank with a minimal match of $0.97$ in between $1-400 M_{\odot}$ that contains at least $10^{6}$ number of template points. Therefore, the overall cost of matched filtering becomes $10^{6} \times N_{s} \log N_{s}$ floating-point operations. With the improvement in the sensitivity of the existing $\GW$ detectors and the inclusion of new detectors such as $\LIGO$-India \cite{saleem2021science} in the global network, matched filtering will become increasingly more expensive.

%%%%------------------------------------------------------------------
Deep learning-based methods, which treat  $\GW$ detection as a classification problem, have the potential to address the issue of latency. Many deep learning-based  searches \cite{krastev2020real, George:2016hay, George:2017pmj, cuoco2020enhancing, lin2020binary, wang2020gravitational, gabbard2018matching, george2018classification, gebhard2019convolutional, li2020some, wei2020gravitational, miller2019effective} 
have been reported that classify $\GW$ signals coming from different $\CBC$ sources. Gabbard et al.\cite{gabbard2018matching} proposed a deep learning network to identify the $\BBH$ signals embedded in the Gaussian noise and compared the performance of their method with matched-filtering operations. Krastev et al. \cite{krastev2020real} extended the deep learning method for real-time detection of binary neutron star ($\BNS$) mergers which are also the source of electromagnetic radiation. A low latency detection and good sky localization of $\GW$ events from $\BNS$ would help quickly alarming the optical telescopes across the globe for real-time follow-up. Huerta et al. \cite{George:2016hay, George:2017pmj, george2018classification} proposed a ``deep filtering'' method to classify and estimate the parameters of $\BBH$ signals via two separate deep learning architectures, one for classification and another for parameter estimation based on regression, respectively.
%%%%--------------------------------------------------------------------
{\color{black}{Deep learning-based methods are rapid but are highly dependent on the efficient training of the architecture using a large data set. Therefore, deep learning architecture can miss an actual $\GW$ signal without efficient training. On the other hand, matched filter-based search scheme is computationally expensive but provides very high confidence in detecting a $\GW$ signal. There is always a trade-off between computational cost and accuracy in detecting signals. Both the methods have merits and demerits to balance this trade-off. Thus, deep learning-based schemes can not replace the matched filtering-based search straightaway. In our proposed scheme, we combine the two approaches for the first time and provide a unique way of reducing the computational cost of the search. This work represents a novel deep learning-based low latency search scheme that not only recognizes a $\BBH$ signal buried in the detector noise but also identifies the probable location in the parameter space promptly where an actual $\GW$ signal lies. However, we do not predict a point estimation of the located signal over the parameter space. {\color{black}{Deep-learning algorithms typically work well and perform robustly in the presence of Gaussian noise at the inputs to the network. However, their performance may vary in an actual search where the noise may also have non-Gaussian components.}} A deep learning-based search pipeline trained and tested using the Gaussian noise may work in the context of real noise as far as patch identification is considered. The same may not work in point estimation as non-Gaussian components of the noise displace the actual parameters. In our method, we first identify the $\BBH$ signal and then identify the patch in the parameter space to which the signal belongs. In this way, we can squeeze the parameter space on which the matched filtering operations are to be performed, thereby reducing the computational cost.}} 

\section{Methodology}
\label{sec:Methodology}
%%%%--------------------------------------------------------------------
This section describes our proposed method by which it is possible to promptly detect a $\GW$ signal by combining the classical method like matched filtering and deep learning scheme like $\CNN$. {\color{black}{This approach first examines the detector output (noisy data) via trained $\CNN$ architecture to know if the data contains any $\GW$ signal.}} If data contains a $\GW$ signal, it is further sub-classified into patches $\mathcal{P}_{1}, \mathcal{P}_{2}, \cdots, \mathcal{P}_{N}$. The patches are defined in the intrinsic parameter space in which the standard search is to be performed using matched filtering scheme. In a standard search for $\GW$ signal from $\CBC$ sources, a $3\text{D}$ aligned spin template bank is used considering three intrinsic parameters: primary and secondary mass components ($m_{1}, m_{2}$) and effective spin parameter $\chi_{eff}$ for the case when spins of two objects are aligned/anti-aligned with orbital angular momentum. A set of analytical waveforms is generated for each template point of the bank. The matched filter operations are carried out between data and analytical waveforms to obtain a set of triggers based on a fixed threshold on $\SNR$. In general, the number of template points is huge for a given parameter space. Therefore, the overall computational cost is increased with the increasing size of the template bank.
In the Advanced $\LIGO$ era, the detector bandwidth and sensitivity are improved; therefore, it is expected that the size of the template bank will increase with several orders and eventually increase the computational cost of performing matched filter operations. However, using $\CNN$ architecture, we can quickly identify the possible region of the template bank in which the signal is located. Thus, we can reduce the parameter space significantly and perform matched filter operations between data and the templates waveforms lying in the patch predicted using $\CNN$. Once the  $\CNN$ architecture has identified the sub-region of the parameter space, we perform the usual matched filter-based search on that specific patch of the search space. The summary of the whole process has been described in Alg-\ref{Alg:CNN-based search} and Fig.\ref{fig:CNN-framework}. 
%%%%--------------------------------------------------------------------

\vspace{1 cm}
%%%%--------------------------------------------------------------------
\begin{algorithm}[H]
\DontPrintSemicolon
\KwIn{$i^{\text{th}}$ data chunk: $d_{i}$}
\KwOut{Set of triggers}
$d_{i} \rightarrow \CNN$ \atcp{First label of classification}
\If{Flag $1 \rightarrow$ data $d_{i}$ contains signal}{Identify the patch $\mathcal{P}_{i}$, where the signal is located. \atcp{Second label of classification}
Compute $\SNR$: $\langle d_{i}, h(t_{i}) \rangle$, where $t_{i} \in \mathcal{P}_{i}$ \atcp{Based on Matched filtering operation}
Fixed a threshold based on $\SNR$ and obtain a set of triggers. 
}
\Else{Change the data chunk and redo the process}
\caption{Combining $\CNN$ and matched filtering operation for quick search of the $\GW$ signal. }
\label{Alg:CNN-based search}
\end{algorithm}

%%%%--------------------------------------------------------------------
\begin{figure}[htbp]
\centering{
\includegraphics[scale = 0.8]{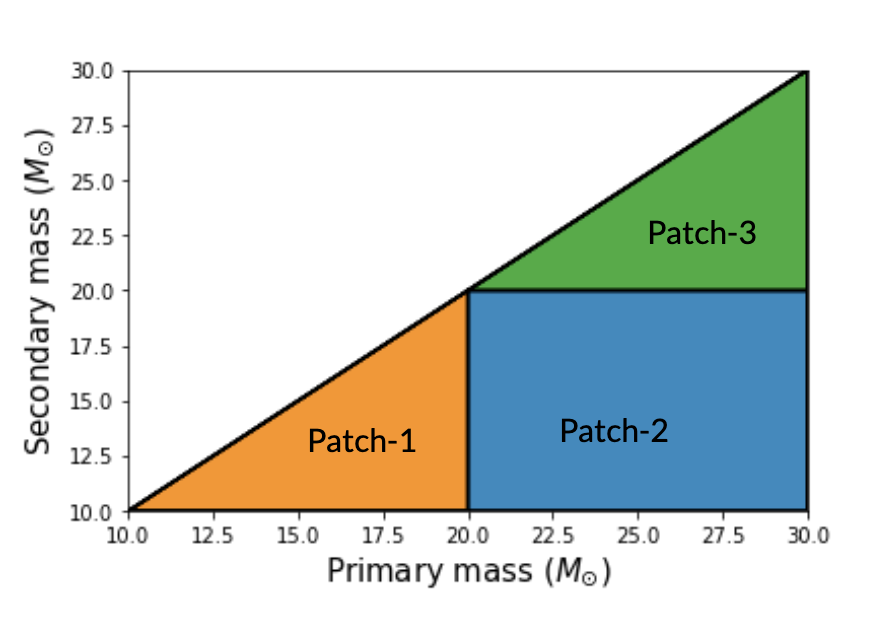}
\caption{
The figure depicts the prescribed methodology for identifying the detected signal in the parameter (intrinsic) space. For simplicity, the depiction has been shown in a two-dimensional space, considering only the primary and secondary mass. However, the method can be further extended to the three or four-dimensional space by adding other intrinsic parameters (e.g., spin components), and patches can be constructed in that space. 
For this example, the figure shows a division of a template bank in between $10-30 M_{\odot}$ into three patches. All the three patches have been constructed in the range $m_{1, 2} : [10-20] \, M_{\odot}$, $m_{1} : [20-30] \, M_{\odot}, m_{2} : [10-20] \, M_{\odot}$, $m_{1,2} : [20-30] M_{\odot}$ respectively considering the parameter space on primary ($m_{1}$) and secondary ($m_{2}$) masses. At the training time of our $\CNN$ architecture, we passed the training data points from these patches individually and labeled them into classes based on their patch id. Therefore, the architecture not only learns about the presence of a $\GW$ signal into the data but further it can learn the possible location in the mass-space. 
It is notable that in the low-latency search ($\GstLAL$ \cite{Cody}) pipeline; the whole bank is divided into sub-banks based on the ranges of chirp-mass ($\mathcal{M}$), $\chi_{\texttt{eff}}$ and duration of the template waveforms. We can also similarly define the patches, and in that case, the patches are nothing but sub-banks based on these parameters. Further, the required number of patches can be user-defined or fixed based on the available computational power. 
}
\label{fig:patch-idea}
}
\end{figure}

\section{$\CNN$ architecture}
\label{sec:architecture}
%%%%-------------------------------------------------------------------
For this work, we adopted a similar $\CNN$ architecture presented in Krastev et al. \cite{krastev2020real}. However, we use a $2\text{D}$ $\CNN$ architecture instead of $1\text{D}$. In general, for the training of time-series data, $1\text{D}$ $\CNN$ architecture is used. We observed that for our data set, $2\text{D}$ $\CNN$ architecture works faster as well as provides better accuracy at the time of training as compared to the $1\text{D}$ $\CNN$ architecture. This is because the number of training parameters is found to be reduced. We have chosen the learning rate of $10^{-4}$, the number of epochs to be $50$, and batch size of $50$. The specific configuration of our architecture, e.g., the number of neurons at each layer, activation function, filter-size, and hidden layers, is shown in Table ~\ref{tab:my_label}. It is notable that for our data set, the configuration used provides an optimal accuracy. However, we believe an alternative configuration can be designed to obtain similar accuracy. 
%%%%--------------------------------------------------------------------
\begin{table}[h!]
\centering{
\begin{tabular}{lccccccc}
\hline
Parameter (Option) &1 &2 &3 &4 &5 &6 &7 \\
\hline
Type& $\text{C}$& $\text{C}$& $\text{C}$& $\text{C}$& $\text{H}$& $\text{H}$& $\text{H}$ \\
{\color{blue}{No. of Neurons}} & $32$ & $64$ & $128$ & $256$ & $128$ & $64$ & No. of classes \\
{\color{blue}{Filter-size}} & ($1$, $16$) & ($1$, $8$) & ($1$, $8$) & ($1$, $8$) & N/A & N/A & N/A \\
{\color{blue}{Max pool size}} & ($1$, $4$) & ($1$, $4$) & ($1$, $4$) & ($1$, $4$) & N/A & N/A & N/A \\
{\color{blue}{Drop out}} & $0$ & $0$ & $0$ & $0$ & $0.5$ & $0.5$ & $0$ \\
{\color{blue}{Activation function}} & $\text{relu}$ & $\text{relu}$ & $\text{relu}$ & $\text{relu}$ & $\text{relu}$ & $\text{relu}$ & $\text{S}_{\text{Max}}$ \\
\hline
\end{tabular}
\caption{The $\CNN$ network consists of four convolution layers ($\text{C}$) followed by two hidden layers ($\text{H}$). Max pooling is performed at each convolution layer. We also use the dropout layer with a rate of $0.5$ at the hidden layers. The last hidden layer has the number of neurons equal to the number of classes used in the experiment. The final layer uses the soft-max ( $\text{S}_{\text{Max}}$) activation function, which gives the output in terms of the prediction probabilities.}
\label{tab:my_label}
}
\end{table}

%%%%--------------------------------------------------------------------
\begin{figure}[h!]
\centering{
\includegraphics[scale = 0.8]{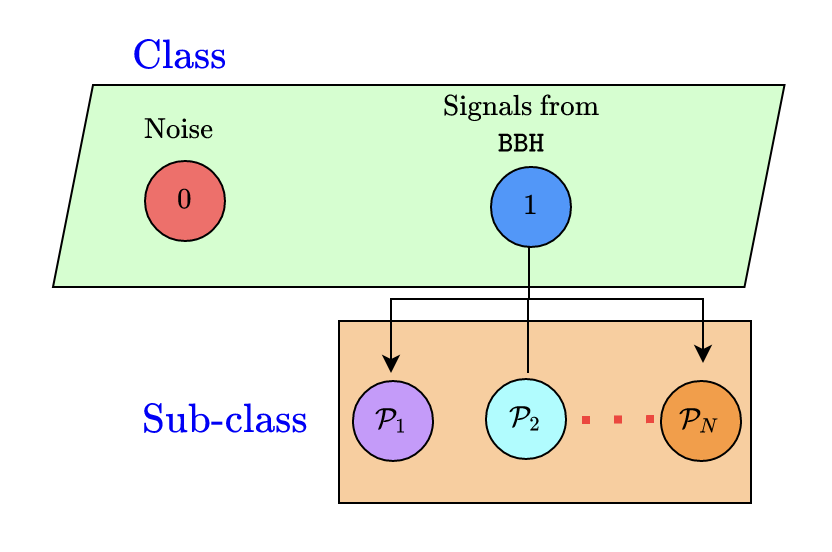}
\caption{ 
The figure demonstrates the $\CNN$ framework proposed in this work. Our proposed architecture used a two-stage classification process to identify the patches $(\mathcal{P}_{1}, \mathcal{P}_{2}, \cdots, \mathcal{P}_{N})$ in which a detected $\GW$ signal can be located. In the first stage, the $\CNN$ architecture classifies the data into two classes: noise (labeled as $0$ in the figure) and the second is $\GW$ signal from $\BBH$ signal (labeled as $1$ in the figure). Once it detected a $\GW$ signal, it can further be classified into patches $\mathcal{P}_{1}, \mathcal{P}_{2}, \cdots, \mathcal{P}_{N}$ in the next label of classification. There are several ways one can define a patch; however, for this work, the defined patches are shown in Fig.~\ref{fig:patch-idea}. 
}
\label{fig:CNN-framework}
}
\end{figure}

\begin{figure}[h!]
\centering{
\includegraphics[scale = 0.7]{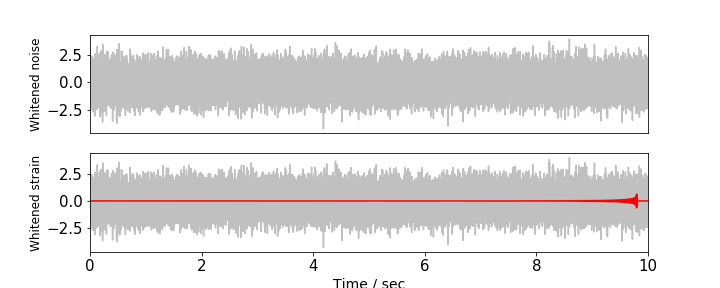}
\caption{ The top panel of the figure shows an example of pure Gaussian noise whitened by the modeled $\PSD$ $\text{aLIGOZerodetHighPower}$, whereas the lower panel shows a whitened strain that contains a $\GW$ signal from $\BBH$ sources. We generated the non-spinning signal of masses $m_{1, 2} = (20, 20)$ using $\IMRPhenomD$ waveform model.   
}
\label{fig:noisy-BBH}
}
\end{figure}

\begin{figure}[h!]
\centering 
\begin{subfigure}{0.49\linewidth}
\centering
\includegraphics[width=\linewidth]{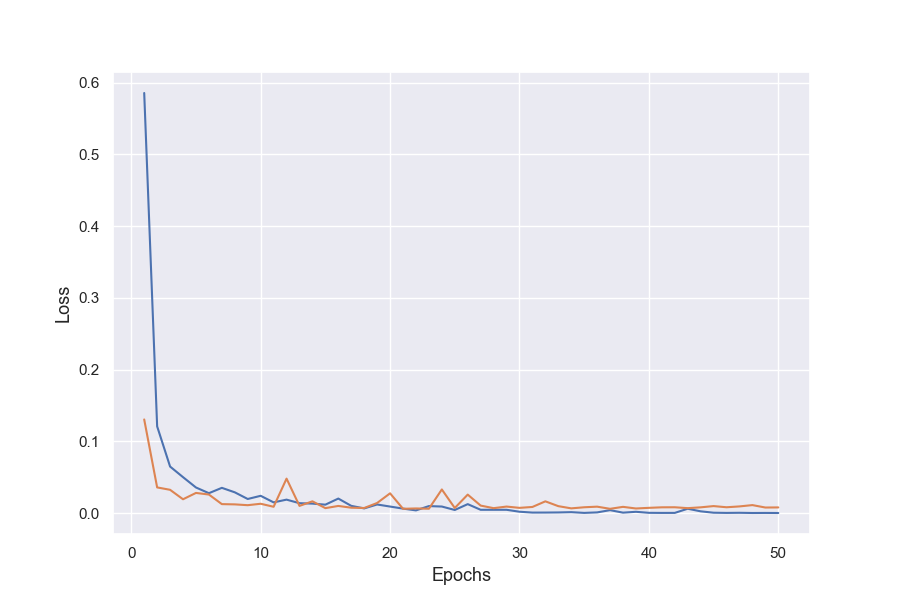}
\caption{Loss}
\label{fig:Loss}
\end{subfigure}
%\hfill
\begin{subfigure}{0.49\linewidth}
\centering
\includegraphics[width=\linewidth]{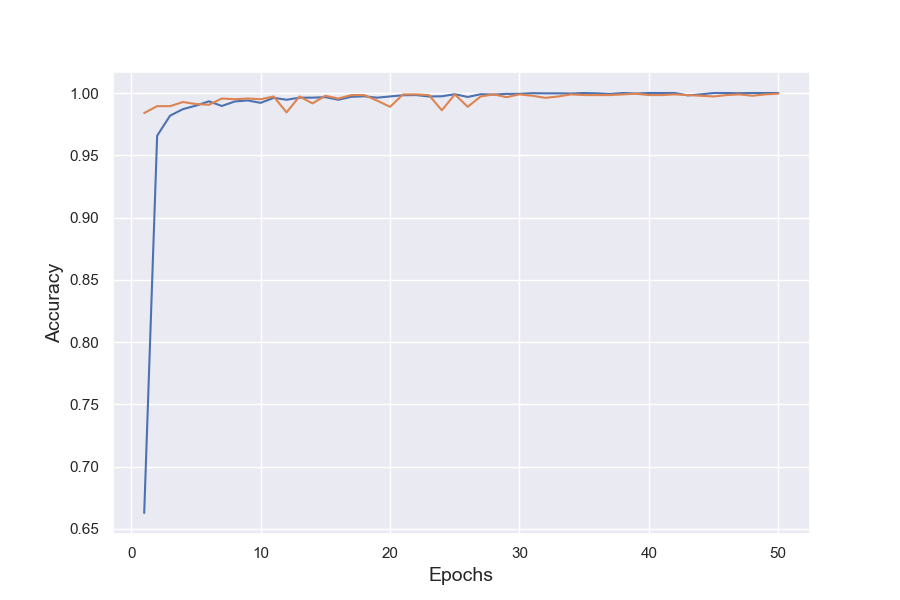}
\caption{Accuracy}
\label{fig:Accuracy}
\end{subfigure}
\caption{The left panel shows the trend of loss in training and validation sets with the varying epochs. Similarly the right panel shows the variation of accuracy in training and validation sets with varying epochs. The same trend in the variation in the loss or accuracy for training and validation indicates the proposed architecture is unbiased.}
\label{fig:Val.loss}
\end{figure}

\section{Analysis Procedure}
\label{sec:analysis procedure}
%%%%--------------------------------------------------------------------
As mentioned earlier, we have framed the whole analysis procedure in two stages. We classify between the $\BBH$ signals (hidden in the Gaussian noise) and the noise at the first level. Fig.\ref{fig:noisy-BBH} shows an example of noise and $\BBH$ signal embedded in noise. We label them as $0$'s and $1$'s, respectively, for our training purpose. The $\BBH$ signals are generated using the $\IMRPhenomD$ model. For the generation of the training data set, the component masses for each binary range from $10 M_{\odot}$ to $70 M_{\odot}$. For simplicity, we have considered non-spinning $\BBH$ systems. We use uniform mass distribution for the generation of the training data set. The extrinsic parameters of each signal, such as right ascension, declination, polarization angle, phase, and inclination angle, are drawn from the uniform distribution. We generate a $10$ sec long Gaussian noise time-series sampled at the frequency $4096 \, \Hz$ and inject the waveforms with their coalescence time fall between $8.0-9.9$ sec. Each simulated signal is whitened by Advanced $\LIGO$'s $\PSD$ at zero-detuned high-power, and the amplitude of each waveform has been re-scaled by optimal $\SNR$.
%%%%--------------------------------------------------------------------
We have chosen the optimal $\SNR$ varying from $10$ to $20$ which is randomly assigned to each signal.
%%%%--------------------------------------------------------------------

%%%%--------------------------------------------------------------------
We classify the $\BBH$ signals at the second level based on their component masses in the mass parameter space. At the time of training, we create a component mass parameter space ranging from $10 M_{\odot}$ to $70 M_{\odot}$ having training data set uniformly distributed across the space. We then divide the whole space equally in three patches having the ranges $(10-30)M_{\odot}$, $(30-50) M_{\odot}$ and $(50-70) M_{\odot}$. We label these three patches as $0$, $1$, and $2$, respectively, for the learning purpose of our architecture. We then train the $\CNN$ architecture to distinguish the $\BBH$ signals from any of these patches. Our architecture gives the probability of a presence of a signal in the detector output, having component mass parameters in one of the defined patches.

\section{Results}
\label{sec:results}
%%%%--------------------------------------------------------------------
In this work, we use TensorFlow \cite{abadi2016tensorflow} based $\CNN$ code base for the training and testing of our designed architecture. Further, we run the whole training process over a single $\GPU$ system. The computational time required for training our architecture for each $\rho_{\text{opt}}$ cases is of $\mathcal{O}(1)$ hour on a single $\GPU$. 
%%%%------------------------------------------------------------------------
\begin{figure}[h!]
\centering
\begin{subfigure}{0.49\textwidth}
\centering
\includegraphics[width=\textwidth]{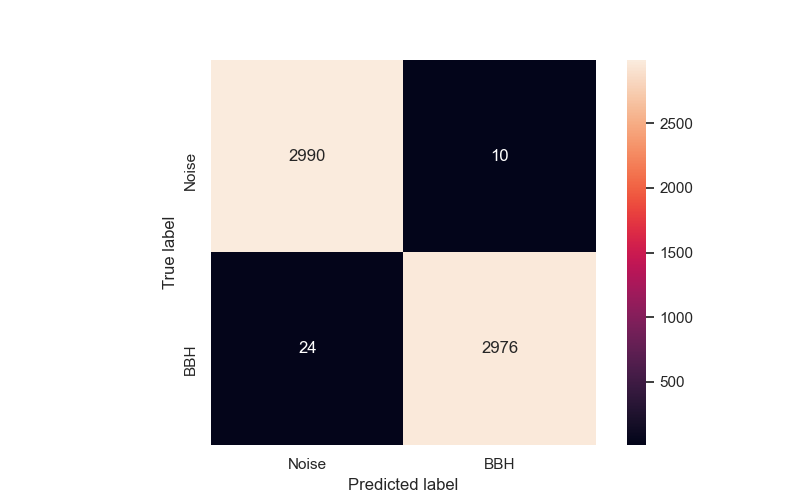}
\end{subfigure}
\begin{subfigure}{0.49\textwidth}
\centering
\includegraphics[width=\textwidth]{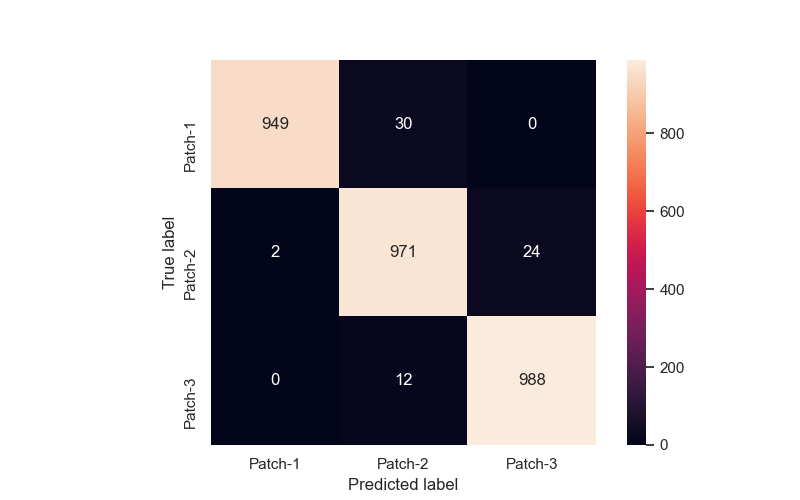}
\end{subfigure}
\caption{The left panel of the figure shows the confusion matrix for the first level of classification between pure noise and noisy $\BBH$ signal.
The right panel of the figure shows the second level of classification for the identification of patches in the mass parameter space. For both figures, the numbers in diagonal squares represent the number of correctly classified signals. Similarly, the numbers in the counter diagonal squares represent the number of miss-classifications. It is evident that for both the cases, the classification accuracy is very high. The first level of classification is around $99\%$, and for the second level overall accuracy is $\geq 97 \%$.
} 
\label{fig:BBH_Noise_CM}
\end{figure}
%%%%--------------------------------------------------------------------
\begin{figure}[h!]
\centering
\includegraphics[scale=0.55]{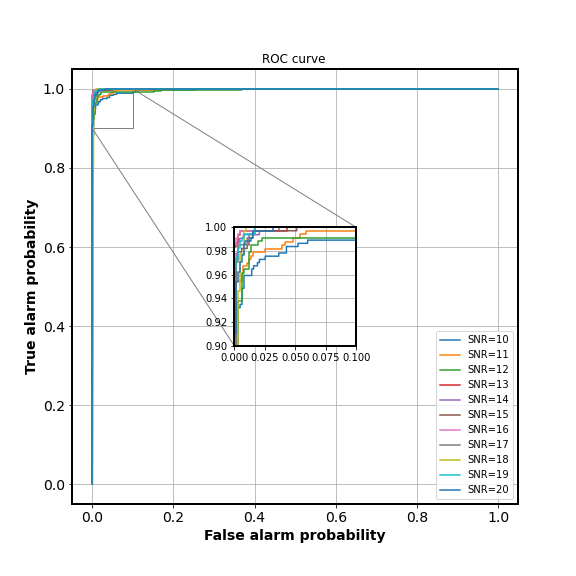}
\caption{
This figure shows the Receiver Operating Characteristics ($\ROC$) curves for each optimal $\SNR$ ($\rho_{opt}$) ranging from $10-20$ shown in different colours for the classification of $\BBH$ signals in each patch. This curve shows the performance of the network improves with the increase in $\rho_{opt}$.}
\label{fig:ROC}
\end{figure}
%%%%-------------------------------------------------------------------
\begin{figure}[h!]
\centering
\begin{subfigure}{0.3\textwidth}
\centering
\includegraphics[width=\textwidth]{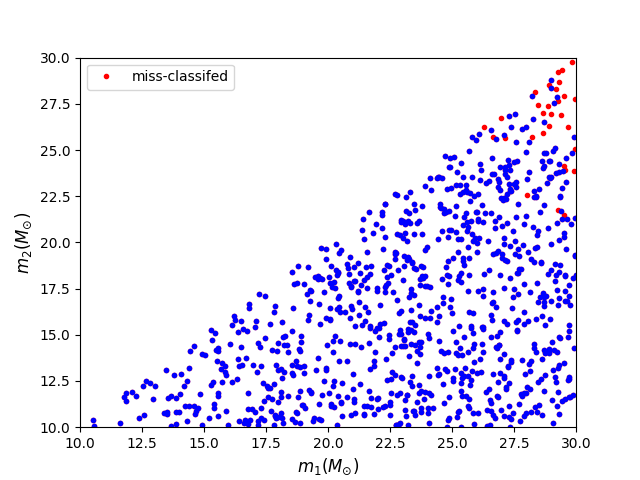}
\caption{Patch-1}
\label{fig:Patch-1}
\end{subfigure}
\begin{subfigure}{0.3\textwidth}
\centering
\includegraphics[width=\textwidth]{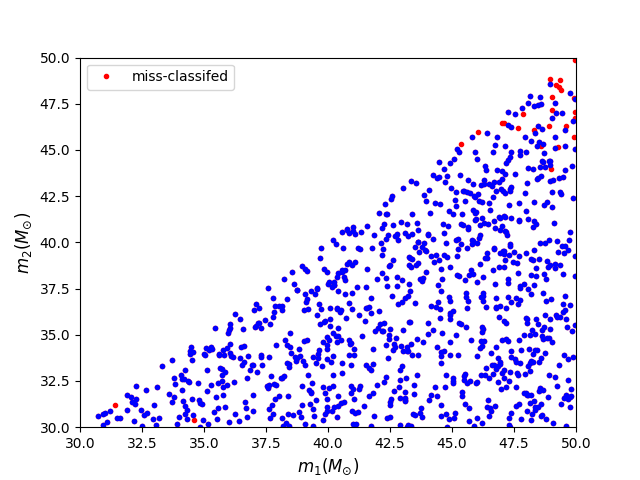}
\caption{Patch-2}
\label{fig:Patch-2}
\end{subfigure}
\begin{subfigure}{0.3\textwidth}
\centering
\includegraphics[width=\textwidth]{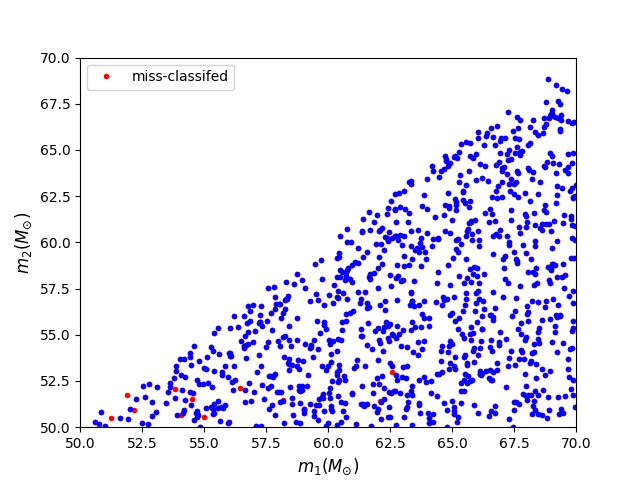}
\caption{Patch-3}
\label{fig:Patch-3}
\end{subfigure}
\caption{In this figure, we show the distribution of correctly classified and mis-classified $\BBH$ signals in blue and red colour respectively for all the three patches.}
\label{fig:Patches}
\end{figure}

We train our $\CNN$ architecture, at the first level, to classify $\GW$ signal from $\BBH$ sources (submerged in Gaussian noise) from pure Gaussian noise (Fig. \ref{fig:noisy-BBH}). We feed $10^{4}$ training samples to the architecture, out of which $50\%$, i.e., $5000$ are pure Gaussian noise samples, and the rest of $50\%$ are noisy $\GW$ signals obtained from $\BBH$ sources. We train the network for $50$ epochs. However, from Fig.~\ref{fig:Val.loss}, it is clear, the maximum accuracy of the network reaches around $10$ epochs only. {\color{black}{That implies that ten epochs are sufficient for optimal training accuracy for our architecture and data set. }}Similarly, at the second level, we train the same architecture to sub-classify $\BBH$ signals into the patches of component masses. We used the same training samples ($10^{4}$) as used in the first level of binary classification. These samples are uniformly distributed across all three patches on component masses. At this level, in order to achieve higher classification accuracy, we employed the curriculum learning \cite{CL} method for the training. To implement curriculum learning, we first train the network with the set of $10^{4}$ loudest (high optimal $\SNR$) $\BBH$ signal of $\rho_{\text{opt}} = 20$ and then decrease optimal $\SNR$ in a step of one up to $\rho_{\text{opt}} = 10$. The performance of the architecture corresponding to each $\rho_{\text{opt}}$ is shown in Fig.~\ref{fig:ROC} in terms of $\ROC$ curves.
%%%%------------------------------------------------------------------------

Once the architecture is trained at both levels, we test it by feeding an independent testing data set of $6,000$ samples equally divided into noisy $\BBH$ signals and pure noise. The architecture at the first level classifies noisy $\BBH$ signal and pure noise with the accuracy of 99\%. The performance of the network at this level is shown in Fig.~\ref{fig:BBH_Noise_CM} (left panel). The correctly classified $\BBH$ signals are further sub-classified into the patches at the second level with the overall accuracy $\geq 97\%$. The network identifies the $\BBH$ signal from the individual patches with the accuracy of $96.9\%$, $97.3\%$, and $98.8\%$ for the first, second, and third patches, respectively, as shown in the right panel of Fig.~\ref{fig:BBH_Noise_CM}. The correctly classified and miss-classified $\BBH$ signals in the patches are shown with blue and red colors respectively in the sub-plot of Fig.~\ref{fig:Patches}. We observe that most of the miss-classified $\BBH$ signals are located near the boundaries of the patches shown in red color in the same figure. This is because the waveform parameters at the boundaries do not differ too much near the interface of the adjacent patches. Hence, architecture finds it challenging to differentiate between the waveforms. To address the issue, we need to pick more training data sets at the adjacency of the patches. In this work, we have chosen three patches only. However, we can perform the same exercise by increasing the number of patches (i.e., defining smaller patches). After obtaining the patches, we need to perform matched filter operation based on the ranked patches. In this work, we have not performed matched filter and other statistical measures which are used in the search pipeline ($\PyCBC$ \cite{usman2016pycbc}, $\GstLAL$). The work is underway and will be reported soon. 
%%%%------------------------------------------------------------------------

%%%%------------------------------------------------------------------------

%%%%--------------------------------------------------------------------

\section{Conclusion}
%%%%--------------------------------------------------------------------
{\color{black}{This work presented a novel strategy to detect $\GW$ signals from $\CBC$ sources}}, exploiting a trained $\CNN$ architecture and matching filter-based trigger identification framework. We use a sample of synthetic data considering a single $\GW$ interferometer for our classification purpose. We explored different choices for the network architecture and finally came up with an optimal configuration that provides binary classification between pure noise and noisy $\BBH$ signal in patches with high accuracy. The proposed strategy comes with a remarkable advantage: under the assumption that if the number of identified patches for a real $\GW$ is significantly less, then the overall computational cost of performing a search will be reduced significantly. Our method can be used to develop a hybrid low-latency search pipeline combining deep learning and matched filtering based on existing search schemes. 
%%%%--------------------------------------------------------------------

We have chosen a simplified setup to test our proposed method in this work, and the initial results are very promising. We want to improve the whole procedure by considering a more realistic setup in which we need to extend our patch identification method using $\CNN$ incorporating the following ideas. \\
%%%%--------------------------------------------------------------------
(a) {\color{blue}{Patch identification in three-dimensional space}}: We want to extend our method for the three-dimensional space by incorporating the ranges of the individual spin components along with mass components. Further, we need to investigate how optimally we can define patches in the three-dimensional space. \\
%%%%--------------------------------------------------------------------
(b) {\color{blue}{Multi-detector-based patch identification}}: For this work, we showed the performance of our method based on one detector only; however, we need to explore the same methodology by adding more detectors. Incorporating the extra detector will help us reduce the false identification of the patches as if there are actual $\GW$ signal presence in both detector data, it can be located to the nearby patches. Therefore, extending this work for the multi-detector case is crucial for obtaining the signal with high confidence. \\
%%%%--------------------------------------------------------------------
(c) {\color{blue}{Exploration for other $\CBC$ sources}}: The proposed method could be more suitable for searching $\GW$ signal from the other $\CBC$ sources like $\BNS$ and $\NSBH$. For these sources, the waveform duration is very long; hence the matched filtering cost is also high. We can explore our method for these sources to reduce the cost. \\
%%%%------------------------------------------------------------------
(d) {\color{blue}{Adaption of the proposed method in low-latency search pipeline}}: We want to map the patch-based covering of the parameter space to the sub-bank-based representation of an aligned template bank used in a low-latency search pipeline ($\GstLAL$). Successful mapping of the patches to the sub-bank will help to implement our idea directly to the $\GstLAL$ search pipeline to reduce the computational cost for the low-latency search scheme. 
%%%%--------------------------------------------------------------------

\begin{acknowledgments}
C.V would like to acknowledge the Council of Scientific \& Industrial Research ($\texttt{CSIR}$), India, for providing financial assistance. S.C and A.R are supported by the Netherlands Organisation for Scientific Research ($\texttt{NWO}$) research program. Also, S.C and A.R are grateful for the computational resources provided by the $\LIGO$ Laboratory. 
\end{acknowledgments}
%%%%--------------------------------------------------------------------
%\nocite{*}
%\bibliography{Reference}
%\bibliographystyle{apsrev4-1}
\bibliography{Reference}

\begin{thebibliography}{10}

\bibitem{aLIGO}
{\sc J.~Aasi et~al.}, {\em Advanced {LIGO}}, Classical and Quantum Gravity, 32
  (2015), p.~074001.

\bibitem{abadi2016tensorflow}
{\sc M.~Abadi, P.~Barham, J.~Chen, Z.~Chen, A.~Davis, J.~Dean, M.~Devin,
  S.~Ghemawat, G.~Irving, M.~Isard, et~al.}, {\em Tensorflow: A system for
  large-scale machine learning}, in 12th $\{$USENIX$\}$ symposium on operating
  systems design and implementation ($\{$OSDI$\}$ 16), 2016, pp.~265--283.

\bibitem{abbott2019gwtc}
{\sc B.~Abbott, R.~Abbott, T.~Abbott, S.~Abraham, F.~Acernese, K.~Ackley,
  C.~Adams, R.~Adhikari, V.~Adya, C.~Affeldt, et~al.}, {\em Gwtc-1: a
  gravitational-wave transient catalog of compact binary mergers observed by
  ligo and virgo during the first and second observing runs}, Physical Review
  X, 9 (2019), p.~031040.

\bibitem{abbott2016gw150914}
{\sc B.~P. Abbott, R.~Abbott, T.~Abbott, M.~Abernathy, F.~Acernese, K.~Ackley,
  C.~Adams, T.~Adams, P.~Addesso, R.~Adhikari, et~al.}, {\em Gw150914: The
  advanced ligo detectors in the era of first discoveries}, Physical review
  letters, 116 (2016), p.~131103.

\bibitem{Abbott:2020niy}
{\sc R.~Abbott et~al.}, {\em {GWTC-2: Compact Binary Coalescences Observed by
  LIGO and Virgo During the First Half of the Third Observing Run}},  (2020).

\bibitem{accadia2011status}
{\sc T.~Accadia, F.~Acernese, F.~Antonucci, P.~Astone, G.~Ballardin, F.~Barone,
  M.~Barsuglia, A.~Basti, T.~S. Bauer, M.~Bebronne, et~al.}, {\em Status of the
  virgo project}, Classical and Quantum Gravity, 28 (2011), p.~114002.

\bibitem{Akutsu:2020his}
{\sc T.~Akutsu et~al.}, {\em {Overview of KAGRA: Detector design and
  construction history}},  (2020).

\bibitem{Findchirp}
{\sc B.~Allen, W.~G. Anderson, P.~R. Brady, D.~A. Brown, and J.~D.~E.
  Creighton}, {\em Findchirp: An algorithm for detection of gravitational waves
  from inspiraling compact binaries}, Phys. Rev. D, 85 (2012), p.~122006.

\bibitem{CL}
{\sc Y.~Bengio, J.~Louradour, R.~Collobert, and J.~Weston}, {\em Curriculum
  learning}, in Proceedings of the 26th Annual International Conference on
  Machine Learning, ICML '09, New York, NY, USA, 2009, Association for
  Computing Machinery, p.~41–48.

\bibitem{cuoco2020enhancing}
{\sc E.~Cuoco, J.~Powell, M.~Cavagli{\`a}, K.~Ackley, M.~Bejger, C.~Chatterjee,
  M.~Coughlin, S.~Coughlin, P.~Easter, R.~Essick, et~al.}, {\em Enhancing
  gravitational-wave science with machine learning}, Machine Learning: Science
  and Technology, 2 (2020), p.~011002.

\bibitem{gabbard2018matching}
{\sc H.~Gabbard, M.~Williams, F.~Hayes, and C.~Messenger}, {\em Matching
  matched filtering with deep networks for gravitational-wave astronomy},
  Physical review letters, 120 (2018), p.~141103.

\bibitem{gebhard2019convolutional}
{\sc T.~D. Gebhard, N.~Kilbertus, I.~Harry, and B.~Sch{\"o}lkopf}, {\em
  Convolutional neural networks: A magic bullet for gravitational-wave
  detection?}, Physical Review D, 100 (2019), p.~063015.

\bibitem{George:2017pmj}
{\sc D.~George and E.~Huerta}, {\em {Deep Learning for Real-time Gravitational
  Wave Detection and Parameter Estimation: Results with Advanced LIGO Data}},
  Phys. Lett. B, 778 (2018), pp.~64--70.

\bibitem{George:2016hay}
\leavevmode\vrule height 2pt depth -1.6pt width 23pt, {\em {Deep Neural
  Networks to Enable Real-time Multimessenger Astrophysics}}, Phys. Rev. D, 97
  (2018), p.~044039.

\bibitem{george2018classification}
{\sc D.~George, H.~Shen, and E.~Huerta}, {\em Classification and unsupervised
  clustering of ligo data with deep transfer learning}, Physical Review D, 97
  (2018), p.~101501.

\bibitem{harry2009stochastic}
{\sc I.~W. Harry, B.~Allen, and B.~Sathyaprakash}, {\em Stochastic template
  placement algorithm for gravitational wave data analysis}, Physical Review D,
  80 (2009), p.~104014.

\bibitem{helstrom2013statistical}
{\sc C.~Helstrom, D.~Fry, L.~Costrell, and K.~Kandiah}, {\em Statistical Theory
  of Signal Detection: International Series of Monographs in Electronics and
  Instrumentation}, International series of monographs on electronics and
  instrumentation, Elsevier Science, 2013.

\bibitem{krastev2020real}
{\sc P.~G. Krastev}, {\em Real-time detection of gravitational waves from
  binary neutron stars using artificial neural networks}, Physics Letters B,
  803 (2020), p.~135330.

\bibitem{li2020some}
{\sc X.-R. Li, W.-L. Yu, X.-L. Fan, and G.~J. Babu}, {\em Some optimizations on
  detecting gravitational wave using convolutional neural network}, Frontiers
  of Physics, 15 (2020), pp.~1--11.

\bibitem{lin2020binary}
{\sc B.-J. Lin, X.-R. Li, and W.-L. Yu}, {\em Binary neutron stars
  gravitational wave detection based on wavelet packet analysis and
  convolutional neural networks}, Frontiers of Physics, 15 (2020), p.~24602.

\bibitem{Cody}
{\sc C.~{Messick}, K.~{Blackburn}, P.~{Brady}, P.~{Brockill}, K.~{Cannon},
  R.~{Cariou}, S.~{Caudill}, S.~J. {Chamberlin}, J.~D.~E. {Creighton},
  R.~{Everett}, C.~{Hanna}, D.~{Keppel}, R.~N. {Lang}, T.~G.~F. {Li},
  D.~{Meacher}, A.~{Nielsen}, C.~{Pankow}, S.~{Privitera}, H.~{Qi},
  S.~{Sachdev}, L.~{Sadeghian}, L.~{Singer}, E.~G. {Thomas}, L.~{Wade},
  M.~{Wade}, A.~{Weinstein}, and K.~{Wiesner}}, {\em {Analysis framework for
  the prompt discovery of compact binary mergers in gravitational-wave data}},
  "Phys. Rev. D", 95 (2017), p.~042001.

\bibitem{miller2019effective}
{\sc A.~L. Miller, P.~Astone, S.~D’Antonio, S.~Frasca, G.~Intini, I.~La~Rosa,
  P.~Leaci, S.~Mastrogiovanni, F.~Muciaccia, A.~Mitidis, et~al.}, {\em How
  effective is machine learning to detect long transient gravitational waves
  from neutron stars in a real search?}, Physical Review D, 100 (2019),
  p.~062005.

\bibitem{owen1996search}
{\sc B.~J. Owen}, {\em Search templates for gravitational waves from
  inspiraling binaries: Choice of template spacing}, Physical Review D, 53
  (1996), p.~6749.

\bibitem{Satya}
{\sc B.~J. Owen and B.~S. Sathyaprakash}, {\em Matched filtering of
  gravitational waves from inspiraling compact binaries: Computational cost and
  template placement}, Phys. Rev. D, 60 (1999), p.~022002.

\bibitem{saleem2021science}
{\sc M.~Saleem, J.~Rana, V.~Gayathri, A.~Vijaykumar, S.~Goyal, S.~Sachdev,
  J.~Suresh, S.~Sudhagar, A.~Mukherjee, G.~Gaur, B.~Sathyaprakash, A.~Pai,
  R.~X. Adhikari, P.~Ajith, and S.~Bose}, {\em The science case for
  ligo-india}, 2021.

\bibitem{usman2016pycbc}
{\sc S.~A. Usman, A.~H. Nitz, I.~W. Harry, C.~M. Biwer, D.~A. Brown, M.~Cabero,
  C.~D. Capano, T.~Dal~Canton, T.~Dent, S.~Fairhurst, et~al.}, {\em The pycbc
  search for gravitational waves from compact binary coalescence}, Classical
  and Quantum Gravity, 33 (2016), p.~215004.

\bibitem{visa2011confusion}
{\sc S.~Visa, B.~Ramsay, A.~L. Ralescu, and E.~Van Der~Knaap}, {\em Confusion
  matrix-based feature selection.}, MAICS, 710 (2011), pp.~120--127.

\bibitem{wang2020gravitational}
{\sc H.~Wang, S.~Wu, Z.~Cao, X.~Liu, and J.-Y. Zhu}, {\em Gravitational-wave
  signal recognition of ligo data by deep learning}, Physical Review D, 101
  (2020), p.~104003.

\bibitem{wei2020gravitational}
{\sc W.~Wei and E.~Huerta}, {\em Gravitational wave denoising of binary black
  hole mergers with deep learning}, Physics Letters B, 800 (2020), p.~135081.

\end{thebibliography}
%%%%--------------------------------------------------------------------

\end{document}